\documentclass[aps,prl,twocolumn,superscriptaddress]{revtex4-2}
\usepackage{graphicx}
\usepackage{xcolor}
\usepackage{hyperref}
\usepackage[all]{hypcap} 
\begin{document}
\title 
{Betweenness centrality illuminates intermittent frictional dynamics}

\author{Omid Dorostkar 
}
\email[]{Correspondence: Omid Dorostkar (domid@ethz.ch) and Dominik Strebel (strebdom@ethz.ch)}
\affiliation{Department of Mechanical and Process Engineering, ETH Z\"urich, Switzerland}
\affiliation{Department of Engineering Science, University of Oxford, Parks Road, Oxford}
\author{Karen E. Daniels 
}
\affiliation{Department of Physics, North Carolina State University, Raleigh, North Carolina, USA}
\author{Dominik Strebel 
}
\email[]{Correspondence: Omid Dorostkar (domid@ethz.ch) and Dominik Strebel (strebdom@ethz.ch)}
\affiliation{Department of Mechanical and Process Engineering, ETH Z\"urich, Switzerland}
\author{Jan Carmeliet}
\affiliation{Department of Mechanical and Process Engineering, ETH Z\"urich, Switzerland}

\date{\today}

\begin{abstract}
Dense granular systems subjected to an imposed shear stress undergo stick-slip dynamics with systematic patterns of dilation-compaction. During each stick phase, as the frictional strength builds up, the granular system dilates to accommodate shear strain, developing stronger force networks. During each slip event, when the stored energy is released, particles experience large rearrangements and the granular network can significantly change. Here, we use numerical simulations of 3D, sheared frictional packings to show that the mean betweenness centrality -- a property of network of interparticle connections -- follows consistent patterns during the stick-slip dynamics, showing sharp spikes at each slip event. We identify the source of this behavior as arising from the connectivity and contact arrangements of granular network during dilation-compaction cycles, and find that a lower potential for connection between particles leads to an increase of mean betweenness centrality in the system. Furthermore, we show that at high confinements, few particles lose contact during slip events, leading to a smaller change in granular connectivity and betweenness centrality.

\end{abstract}

\maketitle
\section{Introduction}

The macroscopic response of a granular system under any external loading originates from grain-scale interactions \citep{RN1157, RN1156, RN122, RN1101}. When an external load is applied to granular materials, particles form contacts and create force chains to sustain and transfer the applied load. There has been extensive work addressing the formation of granular networks and seeking appropriate local, grain-scale metrics which aid the estimation and prediction of the system's global behaviour \citep[\textit{e.g.}][]{RN46, RN1135, RN775}. Traditional studies of the rheology of frictional granular media have used discrete or continuum models using various particle-scale or bulk-scale metrics, but these are not able to account for the multiscale, complex organization of grains and contact forces \citep{RN1119}. 

In recent years, network science has introduced new tools and methods to analyse and probe systems with heterogeneous patterns. This includes approaches accounting explicitly for mesoscale structures, successfully applied to study the physics of granular materials across multiple spatial and temporal scales \citep{RN1161, RN1119}. The use of tools from network science in granular media allows us to develop insight into regular or lattice-like interactions \citep[\textit{e.g.}][]{RN1117, RN1133, RN1136, RN1119, RN1142}. For example, failures in a lattice-like system under compressive and tensile loading are shown to take place mainly in locations with larger geodesic edge betweenness centrality than the mean one in the structure~\citep{RN1118}. Particle betweenness centrality is also suggested as a predictor for forces in 2D granular packings, where the total pressure on each particle in the system correlates to its betweenness centrality value extracted from the geometric contact network~\citep{RN1120}. 

Frictional instabilities in sheared amorphous systems that appear in form of stick-slip dynamics are manifestations of a sudden transition from a solid- to a fluid-like state, usually accompanied with abrupt release of energy ~\citep{RN116, RN70}. In the solid-like state or stick phase, there are few particle rearrangements within the system, and the granular medium is almost jammed under the imposed shear stress. In the fluid-like state, particles can rearrange and the system is at least partially unjammed~\citep{RN1146,RN116, RN1147}. Under shear loading, this transition is accompanied with storage and release of energy; during stick phase the system's frictional strength increases and energy is accumulated in the system. At slip, the stored energy is released through particle rearrangement and fragmentation, leading to energy dissipation~\citep{RN698, RN70, RN1160}. During these cyclic process, with respect to change of volumetric strain, the granular system dilates during the stick phase to accommodate the imposed shear strain, and compacts at slip due to large particle rearrangements~\citep{RN1116, RN1145}. Therefore, even though the system is mechanically considered jammed during the stick phase, there are still small particle rearrangements owing to an overall dilation in the system, which affects the evolution of both the particles' contact network, and the total number of contacts ~\citep{RN660}. 

As for stick-slip dynamics, we hypothesize that the systematic dilation-compaction of a dense granular system is a key factor controlling the granular network architecture and connectivity between particles. Drawing on tools from network science, we examine the contact structure of a sheared granular system undergoing stick-slip dynamics. We track the connectivity of particles and extract grain-scale information to examine  correlations between the evolution of the granular network and the frictional instabilities during stick-slip dynamics. In particular, we look at particle betweenness centrality; this measure is calculated solely on contact status of particles, but depends sensitively on the full network of connections. The questions we address in this work are 1) whether there are systematic patterns in granular network architecture during stick-slip cycles, 2) if there is a relation between the characteristics of frictional instabilities and the properties of granular network and 3) how we can characterize and measure those patterns and relationships. We focus on the temporal and spatial behaviour of granular network during stick-slip dynamics and measure ensemble particle betweenness centrality. We discuss the relationship between particle betweenness centrality and the compaction-dilation cycles, and observe connection between its evolution and the particles' freedom to rearrange into a new configuration. We discuss that in a 3D sheared system, the patterns of force chains are different from the patterns of chains with high particle betweenness centrality. We demonstrate that the time evolution of mean betweenness centrality can be used as an indicator for slip events, and discuss how denser systems' decreased freedom to rearrange affects this manifestation. 

\begin{figure*}[!t]
    \hypertarget{fig_1}{}
    \centering
    \includegraphics[width=16cm]{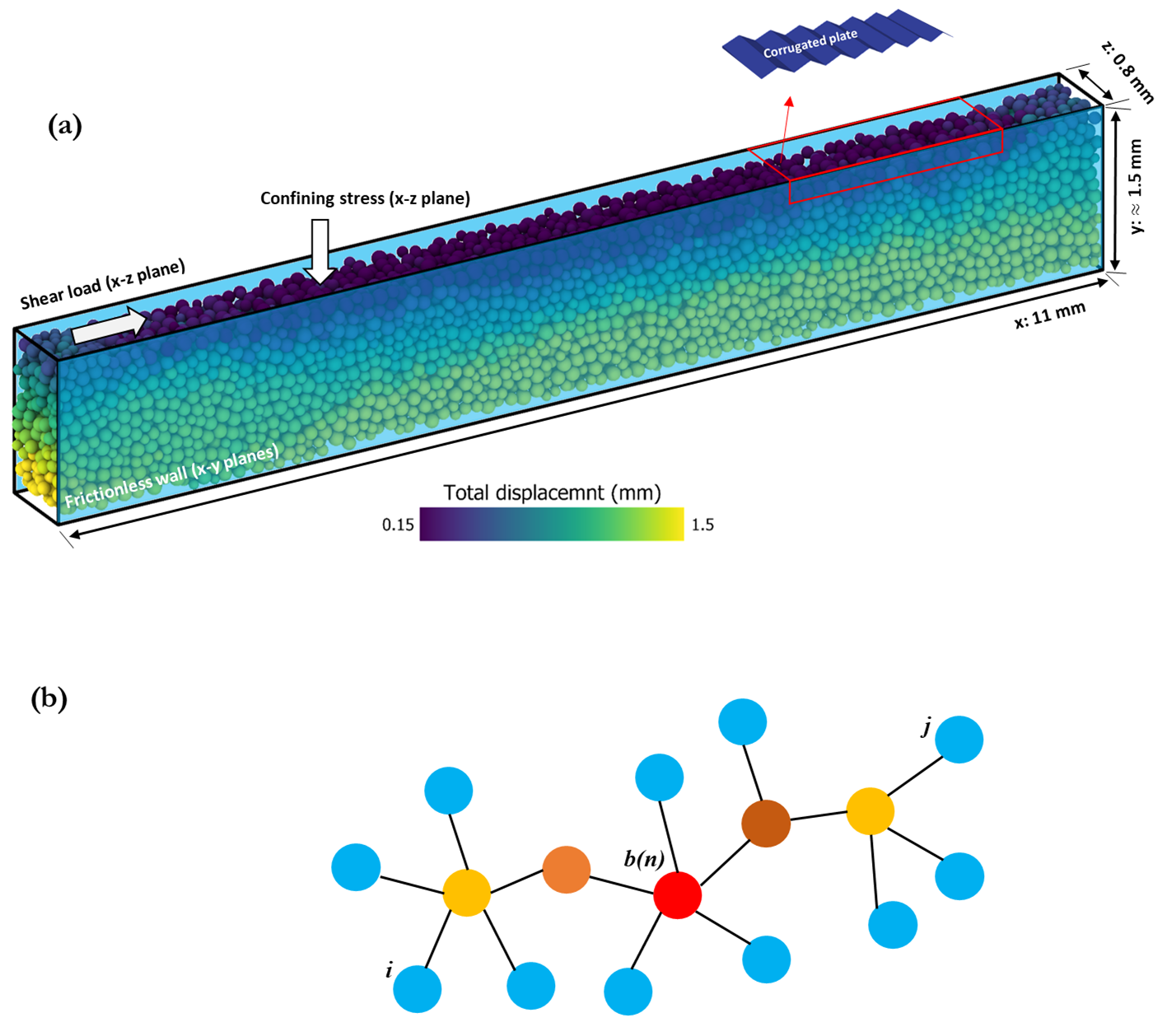}
    \caption{(a) 3-dimensional granular model with frictionless walls at the front and back of the sample, periodic boundaries in $x$-direction, and corrugated plates on top and bottom. (b) Schematic illustration of a connected network of particles with different betweenness centrality values. The red particle has the highest betweenness centrality value, as it is included in many shortest paths between other pairs of particles.}
    \end{figure*}

\begin{figure*}[!t]
    \hypertarget{fig_2}{}
    \centering
    \includegraphics[width=17cm]{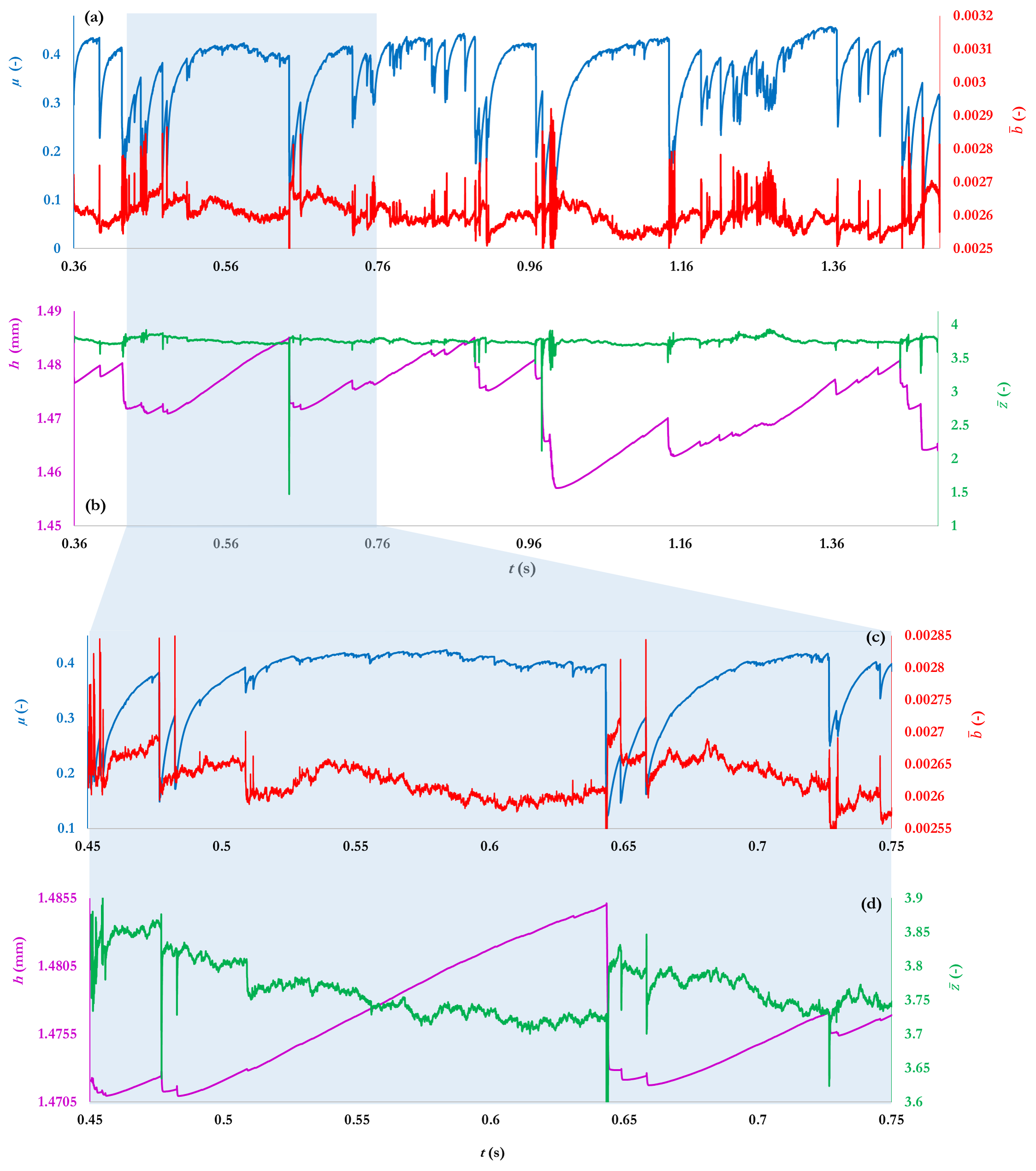}
    \caption{(a) Time series of macroscopic friction $\mu$ and mean betweenness centrality $\overline{b}$ (secondary axis). (b) Time series of sample thickness $h$ (sample size in $y$-direction in Fig.1) and mean coordination number $\overline{z}$. (c-d) Shaded zone from panels (a-b) highlighting the details of several stick-slip cycles. The simulations of this figure are performed under confinement of 500 kPa and shear speed of 0.6 mm/s, and the granular sample has a size of $11\times1.4\times~\mathrm{0.8~mm^3}$.}
\end{figure*}

\begin{figure*}[!t]
    \hypertarget{fig_3}{}
    \centering
    \includegraphics[width=17cm]{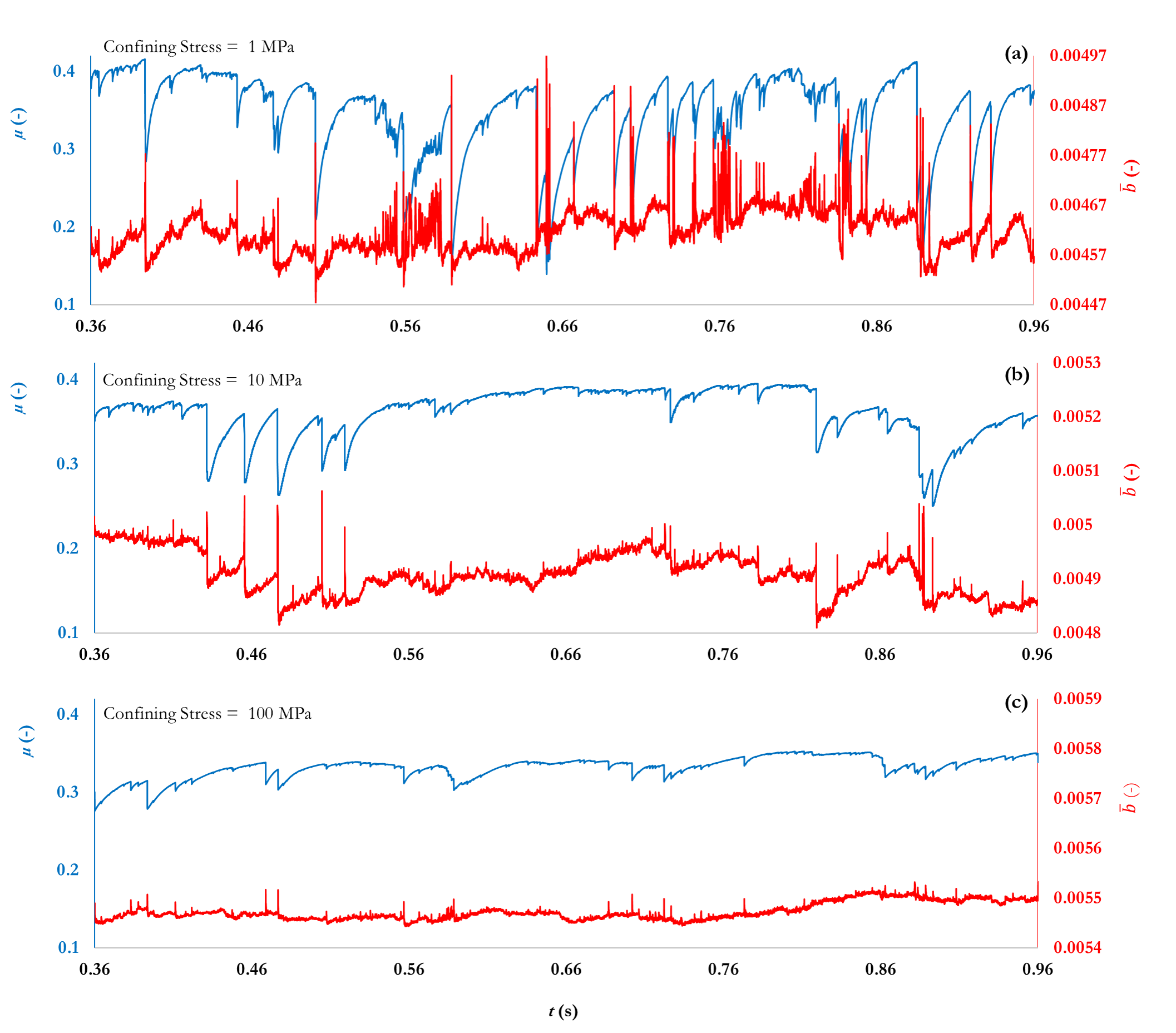}
    \caption{Time series of macroscopic friction $\mu$ and mean betweenness centrality $\overline{b}$ (secondary axis) for confining stresses of 1, 10 and 100 MPa. The simulations of this figure are performed with shear speed of 0.6 mm/s, and the granular sample has a size of $9.2\times1.8\times0.8~\mathrm{mm^3}$.}
\end{figure*}

\begin{figure*}[!t]
    \hypertarget{fig_4}{}
    \centering
    \includegraphics[width=17cm]{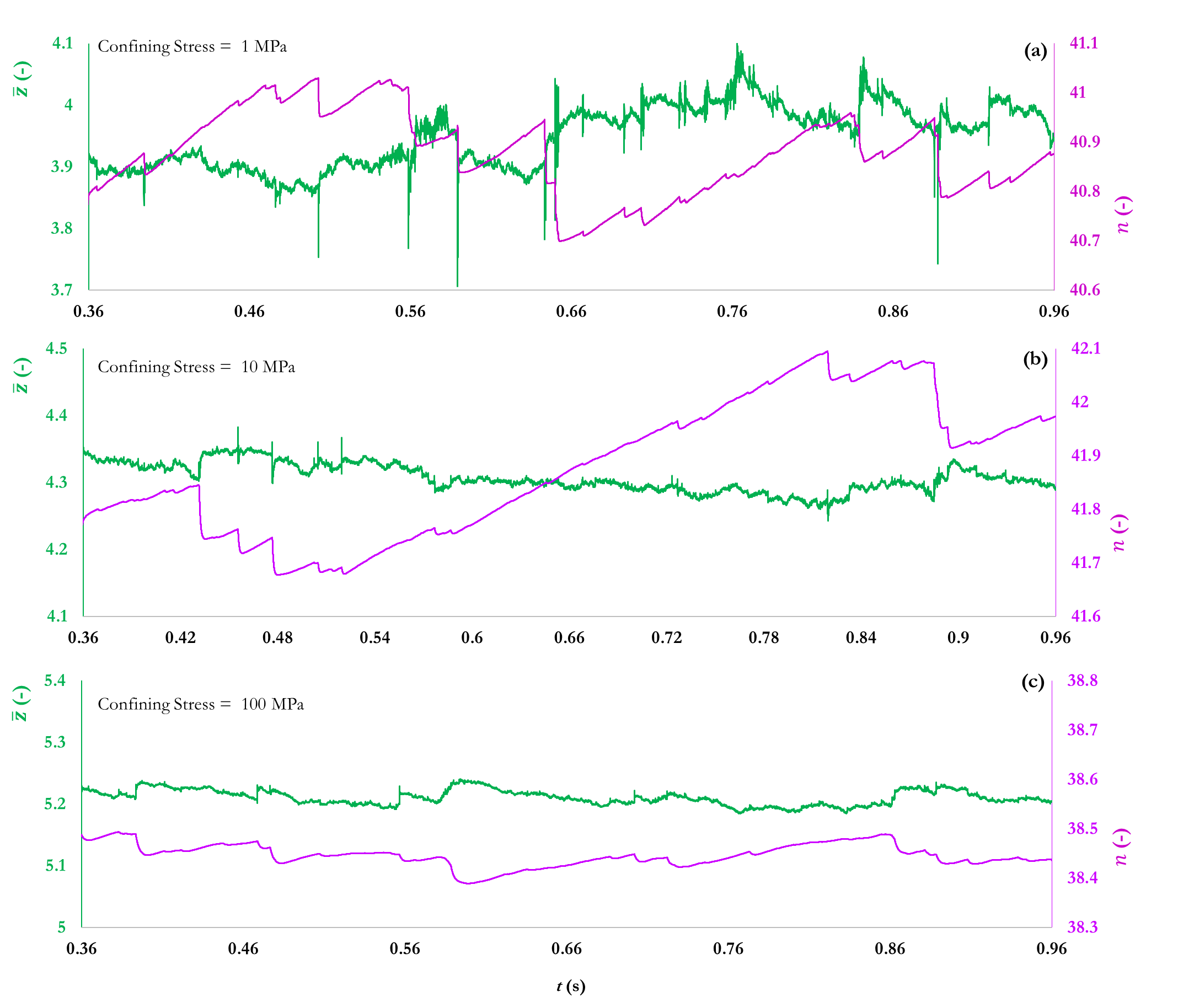}
    \caption{Time series of mean coordination number $\overline{z}$, and sample porosity ${n}$ (secondary axis) for confining stresses of 1, 10 and 100 MPa. The simulations of this figure are performed with shear speed of 0.6 mm/s, and the granular sample has a size of $9.2\times1.8\times~\mathrm{mm^3}$.}
\end{figure*}

\begin{figure*}[!t]
    \hypertarget{fig_5}{}
    \centering
    \includegraphics[width=16cm]{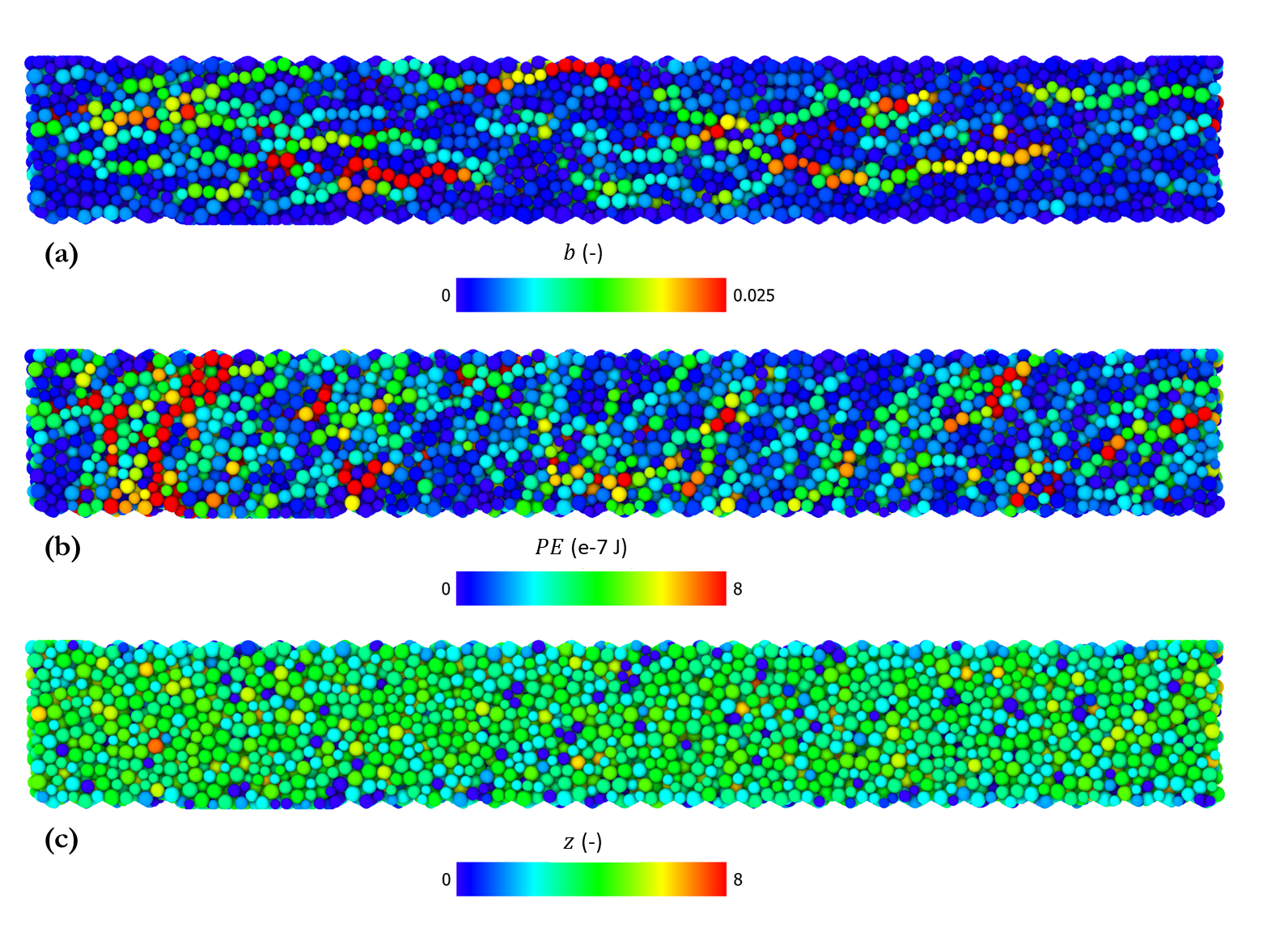}
    \caption{Spatial distribution of (a) particle betweenness centrality $b$, (b) particle potential energy  $PE$ and (c) particle coordination number $z$. The simulations of this figure are performed under confinement of 10 MPa and shear speed of 0.6 mm/s, and the granular sample has a size of $11\times1.4\times0.8~\mathrm{mm^3}$.}
\end{figure*}

\begin{figure*}[!t]
    \hypertarget{fig_6}{}
    \centering
    \includegraphics[width=16cm]{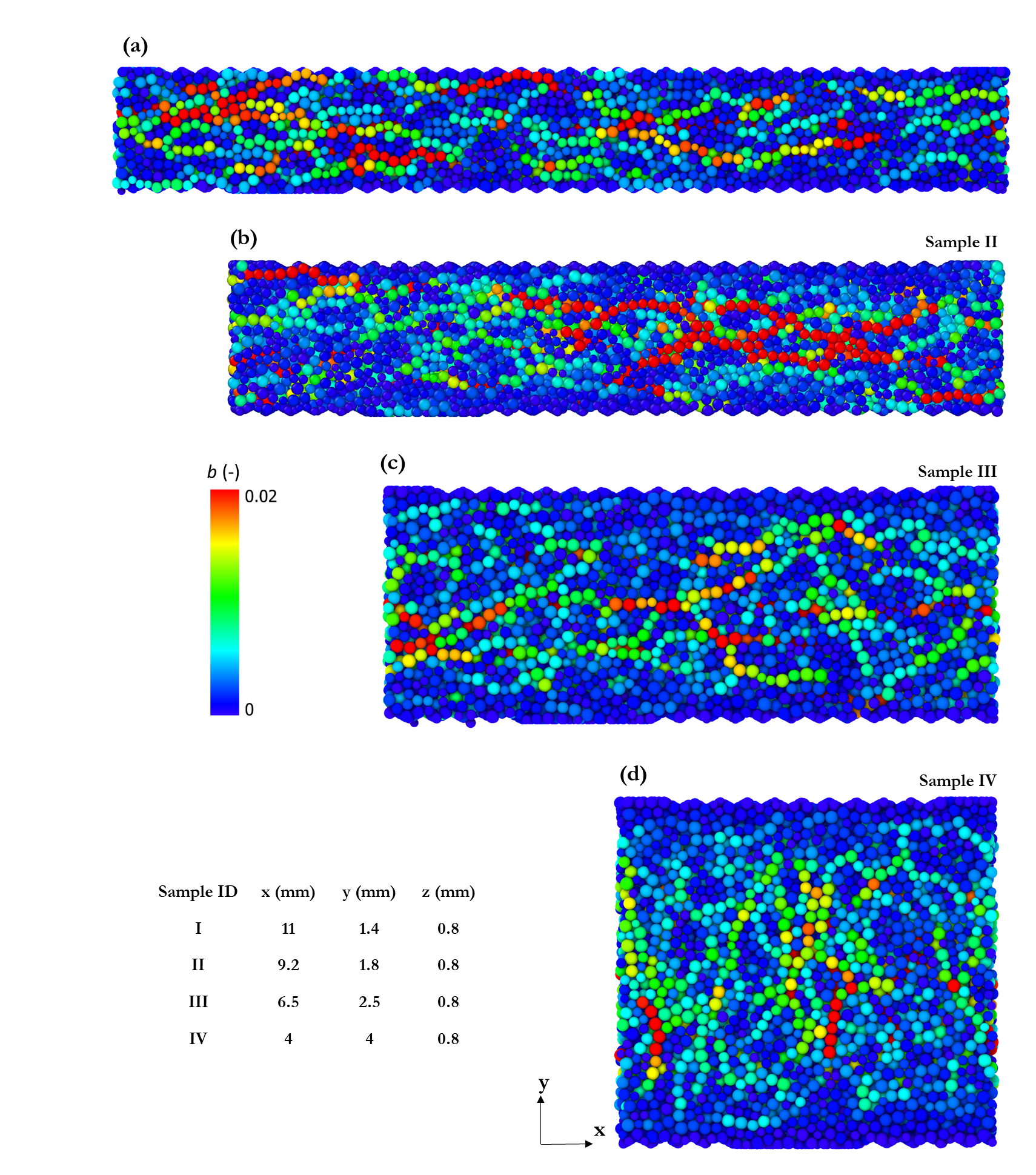}
    \caption{Spatial distribution of particle betweenness centrality b for samples with different measurements in $x$- and $y$-direction but fixed size in $z$-direction, as presented in the legend table. The simulations of this figure are performed under confinement of 10 MPa and shear speed of 0.6 mm/s. The snapshots are taken from the front of the samples and are representative of the whole sample depth ($z$-direction).}
\end{figure*}

\begin{figure*}[!t]
    \hypertarget{fig_7}{}
    \centering
    \includegraphics[width=11cm]{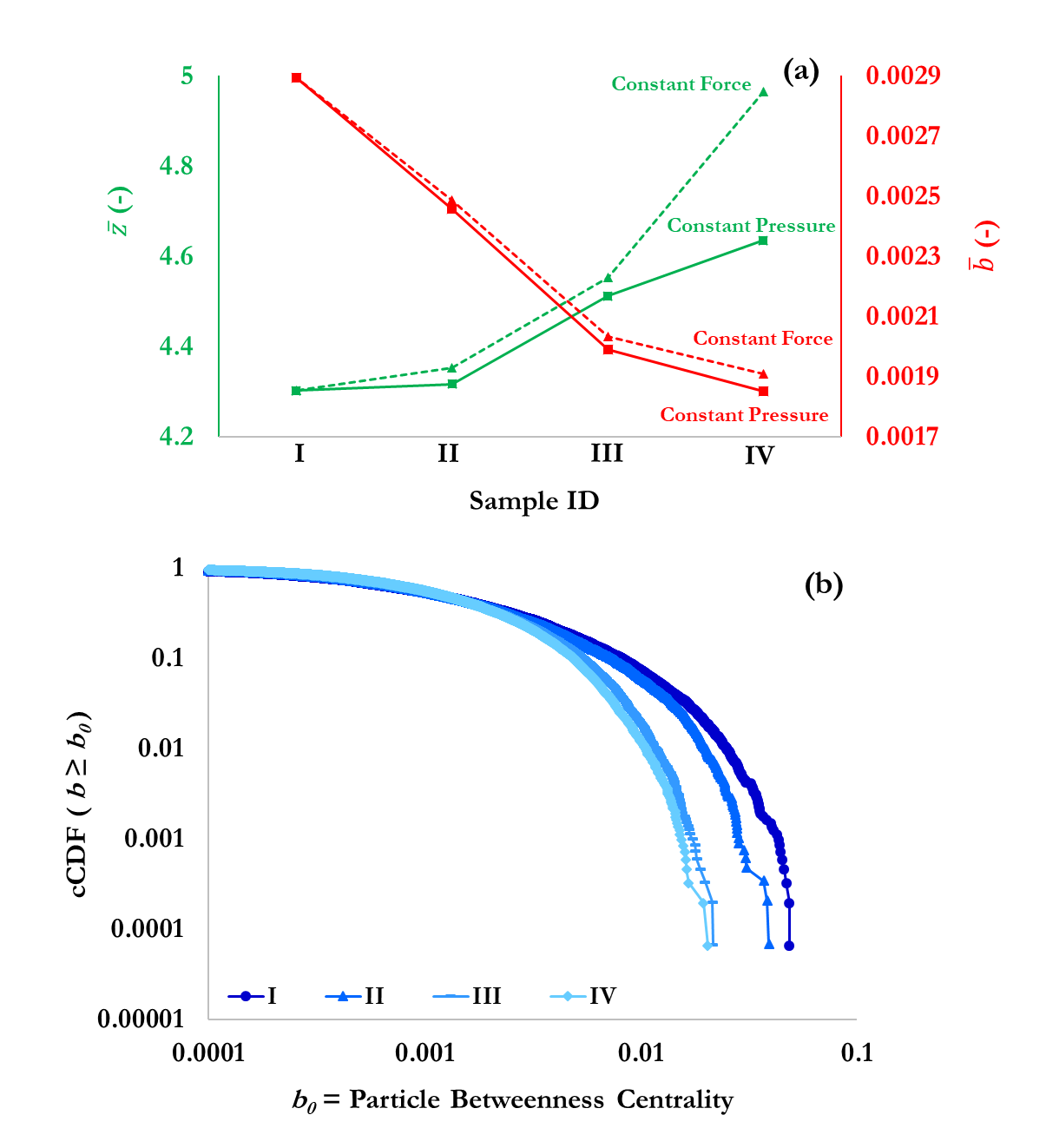}
    \caption{(a) Mean coordination number $\overline{z}$ and mean betweenness centrality $\overline{b}$ (secondary axis) for 4 samples described in \protect\hyperlink{fig_6}{Fig.6}. The original simulations are performed for sample I under confinement of 10 MPa and shear speed of 0.6 mm/s, and the other samples are loaded relative to sample I with either constant pressure (confining stress, shown with continuous line) or constant force (shown with dashed line). (b) Complimentary Cumulative Distribution Function (cCDF) of particle betweenness centrality for 4 points during the stick phase for 4 samples described in \protect\hyperlink{fig_6}{Fig.6}}
\end{figure*}

\section{Method}

We simulate our granular system using standard soft sphere Discrete Element Method (DEM), implemented in the open source software LIGGGHTS \citep{RN41, RN118}. The spherical particles have diameters of $90-150 \: \mathrm{mm}$, drawn from a uniform size distribution; they have a Poisson ratio of $\nu = 0.25$, Young's modulus of $Y = 65 \, \mathrm{GPa}$, restitution coefficient of $r = 0.87$ and inter-particle friction coefficient $\mu_c = 0.5$. The particle density is 2900 $\mathrm{kg/m^{3}}$, leading to a DEM time step of $15 \times 10^{-9} \: \mathrm{s}$.  We apply a constant confining stress via two corrugated plates in the $z$-direction, with a periodical boundary across $x$-direction and frictionless walls in $y$-direction (see \hyperlink{fig_1}{Fig.1a}). The system is sheared using a displacement-control protocol at a constant shear rate of $0.6\:\mathrm{mm/s}$, hence the granular flow is maintained in the quasi-static regime by keeping the inertial number below $10^{-3}$ ~\citep{RN149}. The reference granular model has a sample size of $11\times1.5\times0.8 \:\mathrm{mm^{3}}$ and consists of $N = 7996$ particles; the sample size is changed later to study its effect on betweenness centrality chains, however keeping the number of particles constant. 

In DEM, the equations of motion are solved for each particle:

\begin{equation}
    \sum F_p = m \dot{u_b},
\end{equation}

\begin{equation}
   \sum T_p = I \dot{\omega_b},
\end{equation}

\noindent where $m$, $I$, $u_p$ and $\omega_p$ are the mass, the moment of inertia and the translational and angular velocities of particle, and $F_p$ and $T_p$ are the forces and torques acting on particle, respectively. We use the soft sphere DEM approach, in which the particle-particle contact is modeled with an overlap between them and the contact law is described by a combination of different rheological elements~\citep{RN660}. Using the nonlinear Hertzian particle-particle contact law, the normal and tangential contact forces are described as~\citep{RN193, RN196}: 

\begin{equation}
F_{pn} = -k_{pn} \delta\epsilon_{pn} + c_{pn} \delta u_{pn} ,
\end{equation}

\begin{equation}
F_{pt} = min (|k_{pt} \int_{t_{c,0}}^t \:\delta u_{pt}dt + c_{pt} \delta u_{pt}|, \:  \mu_c F_{pn}) ,
\end{equation}

\noindent where $k_{pn}$ and $k_{pt}$ are the normal and tangential spring stiffness, $c_{pn}$ and $c_{pt}$ are the normal and tangential damping coefficients, $\delta\epsilon_{pn}$ is the overlap, $\delta u_{pn}$ and $\delta u_{pt}$ are the relative normal and tangential velocities, and $\mu_c$ represents the inter-particle friction coefficient, respectively. The normal and tangential spring and damping coefficients are calculated from~\citep{RN193, RN196}:

\begin{equation}
k_{pn} = \frac {4}{3} Y^* \sqrt{R^* \delta\epsilon_{pn}} ,
\end{equation}

\begin{equation}
k_{pt} = 8 G^* \sqrt{R^* \delta\epsilon_{pn}} ,
\end{equation}

\begin{equation}
c_{pn}= \frac{-2 \sqrt{(5/6)}\:ln(r)}{\sqrt{ln^2 (r)+ \pi^2}} \: \sqrt{2Y^* m^* \sqrt{R^*\delta\epsilon_{pn}}}\:, 
\end{equation}
\begin{equation}
c_{pt}= \frac{-2 \sqrt{(5/6)}\:ln(r)}{\sqrt{ln^2 (r)+ \pi^2}} \: \sqrt{8G^* m^* \sqrt{R^*\delta\epsilon_{pn}}}\:, 
\end{equation}

\noindent where $r$ is the restitution coefficient, and $Y^*$, $R^*$, $G^*$ and $m^*$ are the equivalent Young’s modulus as $1/Y^* =  ((1-\nu_1^2))/Y_1 +  ((1-\nu_2^2))/Y_2$, the equivalent radius as $1/R^* =  1/R_1 +  1/R_2$, the equivalent shear modulus  as  $1/G^* =  (2(2-\nu_1)(1+\nu_1))/Y_1 +  (2(2-\nu_2)(1+\nu_2))/Y_2$   and the equivalent mass as $1/m^* =  1/m_1 +  1/m_2$~\citep{RN193, RN196}. The subscripts 1 and 2 refer to the two specific particles in contact and $\nu$ is the Poisson's ratio of the particle. 

We use betweenness centrality $b$ (\hyperlink{fig_1}{Fig.1b}) as a non-local measure of granular network connectivity. The particle betweenness centrality measures the number of times that the shortest paths between a pair of other particles travel through that particle. Mathematically, the betweenness centrality $b$ of particle $n$ is defined as the fraction of shortest paths $S_\mathrm{ij}$ that connect particles $i\neq j \neq n$, going through particle $n$:  

\begin{equation}
    b(n) =\sum\limits_{i\neq j \neq n} \frac{S_\mathrm{ij }(n)}{S_\mathrm{ij}} .
\end{equation}

Since the betweenness centrality of a particle scales with the number of pairs of particles, due to the summation indices, we rescale it by dividing $b(n)$ by a factor of $(N-1)(N-2)/2$, where $N$ is the total number of particles. To calculate particle betweenness centrality, we use open-source functions provided by the Brain Connectivity Toolbox and the Boost Graph Library ~\citep{RN1152, RN1159}. 


\section{Results}
\hyperlink{fig_2}{Fig.2a} shows the time series of the macroscopic friction coefficient $\mu$, calculated as the ratio of shear stress to confining stress, and mean particle betweenness centrality, $\overline{b}$, averaged over the whole sample. The friction signal exhibits irregular stick-slip dynamics, where the slip events have a variety of magnitudes and recurrence times. Prior to slip, the sample is in its critical state; this  preslip period is characterized by deformation being accommodated by microslips mainly due to small particle rearrangements \citep{RN10}. The stick-slip cycles show different critical states in which some cycles experience many microslips. During the stick phase, the $\overline{b}$ signal shows slight variations, gradually increasing during the approach to the eventual slip event. At slip, we consistently observe spikes in $\overline{b}$, and sharp drops after the slip is complete. 

The sample thickness ${h}$ (sample size in $y$-direction) and average coordination number $\overline{z}$ also show consistent patterns following stick-slip dynamics (\hyperlink{fig_1}{Fig.1}). The granular sample dilates during the stick phase and compacts at slip. Due to the dilation during the stick phase, there is a slight decrease in $\overline{z}$. At slip, $\overline{z}$ shows a sharp drop; owing to the compaction of granular layer, $\overline{z}$ recovers and particles again gain more contacts. In the zoomed-in region (see \hyperlink{fig_2}{Fig.2c-d}), we clearly observe this trend of $\overline{b}$ showing gradual increase during stick phase and a sharp spike at slip. The evolution of $\overline{b}$ during the critical state is sensitive to the occurrence of microslips: approaching a major slip event, the increase in $\overline{b}$ is affected by occurrence of microslips, where each microslip slightly reduces $\overline{b}$. For this reason, wherever there are microslips, the gradual increasing trend in $\overline{b}$ during the stick phase stops and the signal instead undergoes many small drops. 

As shown in \hyperlink{fig_2}{Fig.2d}, as the granular layer dilates, the mean coordination number $\overline{z}$ shows a clear decrease during the stick phase, followed by a drop and then recovery at slip. This occurs because, as the shearing advances, the granular system dilates:  due to the dilation and increase of porosity (decrease of packing fraction), some contacts are inevitably lost ~\citep{RN660}. At slip, the granular layer undergoes substantial (as compared to the stick phase) particle rearrangements and ultimately compacts. The drop in $\overline{z}$ shows the comparatively fluid-like (or unjammed) behaviour of granular system, and the recovery of $\overline{z}$ is directly caused by compaction and decrease of porosity during slip~\citep{RN698}. 

It therefore seems that the evolution of $\overline{b}$ is directly related to the behavior of particles that rearrange and change contact status: during the stick phase, as the system dilates and the pore volume expands, some particles lose contacts and the granular contact network is partially lost. Where there is contact loss, the spatial distribution of the granular contact network also becomes less uniform and the connectivity load i.e. the number of shortest paths between particles in the system for the surviving contacts, increases on the remaining contacts, which causes on average an increase of $\overline{b}$. It is important to note that this change in centrality of the remaining connected particles does {\it not} necessarily mean a change in the status of their contacts or their position, but only that the shortest path connecting a pair of other particles can change from a lost path to another one that includes these remaining contacts (remaining connected particles). Betweenness centrality is by definition a measure of the centrality of particles within the ensemble of shortest paths. Therefore, if particles' connectivity and contact network are formed such that some particles get extreme centrality, the betweenness centrality increases on average.As was observed for the behaviour of $\overline{b}$ during the stick phase, and due to large drops in $\overline{z}$ at slip, many connectivity paths are lost and hence the connectivity load increases significantly on a few remaining chains leading to a sharp increase in $\overline{b}$. As the system compacts, the connectivity paths are recovered, the connectivity load is more evenly distributed inside the sample, and $\overline{b}$ drops. An analogy for this behavior is traffic flow between two cities. Suppose there are two main highways connecting two major cities. If highway A gets closed for maintenance, its traffic load will relocate to highway B, which means cars that were already in highway B are now more in the center of the flow (more cars, more shortest paths between a pair of them that include the cars in the middle of the highway), and their betweenness centrality increases. A closure of highway A is similar to the loss of contacts at slip in our model.

We additionally observe that $\overline{b}$ can change by alterations of the confining stress, which controls the connectivity of the granular system. As shown  in \hyperlink{fig_3}{Fig.3}, by increasing the confining stress, the evolution of $\overline{b}$ neither shows a clear gradual increase during the stick phase nor a sharp increase at slip for very high confinements. Particularly at slip events, the amplitudes of the spikes in $\overline{b}$ are very small, and almost indistinguishable from the background values. Another important observation in \hyperlink{fig_3}{Fig.3} is the change in the size of slip friction drops, defined as the change in $\mu$, where at high confinements, lower slip friction drops are observed. Slip friction drop demonstrates the ability of a granular system to release the accumulated shear stress relative to its confinement. A smaller drop in friction for samples with higher confinement (lower porosity) is due to lower potential and freedom for particle rearrangements: during a frictional instability, the system can release smaller amount of stress relative to its confinement, as compared to a weakly confined system. 

In \hyperlink{fig_4}{Fig.4} we demonstrate the evolution of mean coordination number, $\overline{z}$, and porosity, ${n}$, defined as the ratio of pore volume to total volume of the sample. At low confinement, during the stick phase ${n}$ increases as the sample dilates and $\overline{z}$ decreases as the system loses contacts. The sharp drops are clear for $\overline{z}$ at slip events. This consistent behaviour of $\overline{z}$ and ${n}$ are less observable at higher confinements. On the other hand, with increasing confinement, the mean value of ${n}$ and $\overline{z}$ decreases and increases, respectively. The observations in \hyperlink{fig_4}{Fig.4} demonstrate that particles in samples with higher confinement have lower freedom and potential for rearrangement during stick-slip dynamics, supporting our hypothesis as the underlying mechanism for the behaviour of $\overline{b}$ in \hyperlink{fig_3}{Fig.3}. 

For comparison with more traditional measures of granular micromechanics, we consider the spatial distributions of potential energy, betweenness centrality, and coordination number (\hyperlink{fig_5}{Fig.5}). As an approximation, we calculate contact potential energy as $PE= F_{pn}^2/k_{pn} + {F_{pt}^2}/k_{pt}$, and distribute this quantity evenly between the two particles in contact. The total potential energy of a particle is the sum of energies gained from all its contacts. An interesting observation in \hyperlink{fig_5}{Fig.5} is that, unlike the rather uniform distribution observed for coordination number, both particle betweenness centrality and potential energy form chain-like patterns. However, contrary to the energy chains that form diagonally to resist the applied load (and which correspond to conventional notions of force chains \cite{Howell_1999,Kondic_2004}), the betweenness centrality chains show horizontal patterns, aligned with the flow direction. A quantitative study of their properties is beyond the scope of this work, but we will next undertake a qualitative investigation of the betweenness centrality chains structures, which we find to arise from the geometry of the sample.

To investigate the origin of horizontal alignment of betweenness centrality chains in \hyperlink{fig_5}{Fig.5a}, we seek to understand whether they are formed along the direction of applied shear stress i.e. $x$-direction (see \hyperlink{fig_1}{Fig.1}), or are related to smaller dimension of the sample along $y$-direction (\hyperlink{fig_1}{Fig.1}). To shed more light, we perform simulations on granular assemblies with variable aspect ratio: different sizes in $y$-direction but the same number of particles. Therefore, the length of the sample in $x$-direction is smaller for the samples with larger size in $y$-direction (\hyperlink{fig_6}{Fig.6}). Note that the boundary conditions in $x$-direction still remain periodic for all samples, meaning there is no influence of sample length on simulated behavior. Starting with sample I and by increasing the sample size in $y$-direction towards sample IV, the betweenness centrality chains become more diagonal or vertical. This shows that sample size in $y$-direction affects and guides the patterns of betweenness centrality chains, since particles have more possibilities for forming contacts in sample IV. In addition, we also observe that, while sample I shows betweenness centrality chains with some extreme ${b}$ values (red chains), ${b}$ decreases for samples with larger sample size in $y$-direction. The observations in \hyperlink{fig_6}{Fig.6} confirms that if particles are provided with more options for forming contacts, the betweenness centrality chain patterns become more uniformly distributed, even though some extreme chains are present in the sample. This point inferred from the spatial patterns of $b$ is consistent with the mechanism explained for the temporal evolution of $\overline{b}$ during the stick-slip dynamics. As we progress during the stick phase, particles have less options to form new connectivity paths, and therefore extreme $b$ values appear, leading to an increase of $\overline{b}$. This behaviour reaches its ultimate state at occurrence of slip, where because of the loss of many contacts, the connectivity paths are lost, and centrality increases on the remaining chains, leading to a spike in $\overline{b}$.

\hyperlink{fig_7}{Fig.7a} quantitatively confirms that $\overline{b}$ decreases with increasing sample size in $y$-direction, whereas $\overline{z}$ increases. This interesting observation, that with higher number of contacts the mean betweenness centrality is lower, implies that in a sample with larger size in $y$-direction, there are more possibilities for each particle to make the contact with its neighbouring particles, such that the connectivity of particle is more uniformly distributed and $\overline{b}$ is smaller. Note that, with changing sample size in $y$-direction from sample I to IV, we perform the simulations with two loading mechanisms: one at constant pressure (confinement) and the other at constant force, and the observations are valid for both loading protocols (\hyperlink{fig_7}{Fig.7a}). 

We also show in \hyperlink{fig_7}{Fig.7b} the complimentary Cumulative Distributions Functions (cCDF) of ${b}$ for 4 samples described in \hyperlink{fig_6}{Fig.6}. The distributions are made for 4 random points during a stick phase considered to be representative for the whole simulation period. The cCDFs in \hyperlink{fig_7}{Fig.7b} show that, the extreme betweenness centrality chains in \hyperlink{fig_6}{Fig. 6} are caused by only a small portion of particles, as the extreme tails in \hyperlink{fig_7}{Fig.7b} show the deviation of distributions at around 10 percent. We highlight here that our observations for spatial patterns of betweenness centrality chains, and the relation of aspect ratio (different relative sample sizes in $x$- and $y$-direction) with $\overline{b}$ and $\overline{z}$ are consistent with the behaviour of mean betweenness centrality during stick-slip cycles. When particles have more possibilities for making contacts, betweenness centrality is smaller; this situation occurs, for instance, at the beginning of a stick phase or in samples with lager sample size in $y$-direction. On the other hand, when connectivity is limited, whether due to a loss of contacts at slip or to a thinner sample, the centrality experiences higher values. 

\section{Conclusions}

We model stick-slip dynamics in a sheared granular system and study the evolution of network connectivity during frictional intermittent failures using particle betweenness centrality. The mean particle betweenness centrality shows sensitivity to the friction level during stick-slip cycles, controlled by the coordination number and the freedom of particles for rearrangement. In high porosity samples, as occurs at  lower confining pressure due to dilation of the granular system, the mean betweenness centrality increases gradually during the stick phase and spikes at slip. The mean betweenness centrality drops along with the compaction phase following slip. With increasing confinement and decrease of porosity, both the contact loss during the stick phase and the substantial rearrangements during the slip phase become smaller; therefore, the connectivity of network of particles only slightly changes, leading to a less prominent change of mean betweenness centrality. The lower freedom of the  particles to rearrange at very high confinement also limits the granular sample's ability  to release stored shear stress (relative to its confinement) during the frictional instabilities, as the friction drops are smaller at slip. Our results in this work show that betweenness centrality, a metric that solely deals with geometric contact network, can be an indicator for approach of frictional instabilities in sheared granular systems.   

\section{Acknowledgements}

Authors thank Empa for infrastructural supports. KED is grateful for the support of the James S. McDonnell Foundation. 

\section{References}


\end{document}